\documentclass[12pt,fleqn]{article}
\usepackage{amsfonts}
\usepackage{xypic}
\newcommand{\ol}{\setlength{\itemsep}{0pt.}\begin{enumerate}}
\newcommand{\eol}{\end{enumerate}\setlength{\itemsep}{-\parsep}}
\title{\bf Gravity on parallelizable manifold.}
\author{
\thanks {\quad  email kaniel@sunset.ma.huji.ac.il}
Shmuel Kaniel \\
\thanks {\quad   email itin@sunset.ma.huji.ac.il}
Yakov Itin \\
Institute of Mathematics\\
Hebrew University of Jerusalem\\
Givat Ram, Jerusalem 91904, Israel
}
\date{}
\begin{document}
\maketitle

\newtheorem{THEOREM}{Theorem}[section]
\newenvironment{theorem}{\begin{THEOREM} \hspace{-.85em} {\bf :} 
}%
                        {\end{THEOREM}} 

\newtheorem{PROPOSITION}[THEOREM]{Proposition}
\newenvironment{proposition}{\begin{PROPOSITION} \hspace{-.85em} 
{\bf :} }%
                            {\end{PROPOSITION}}


\newcommand{\pro}{\begin{proposition}}

\newcommand{\prf}{\noindent{\bf Proof:} }

\newcommand{\epro}{\end{proposition}}

\newcommand{\eprf}{\bbox}
\newcommand{\beqn}{\begin{equation}}
\newcommand{\eeqn}{\end{equation}}
\newcommand{\bbox}{\vrule height7pt width4pt depth1pt}
\newcommand{\qed}{\bbox}
\
\def\hook{\hbox{\vrule height0pt width4pt depth0.5pt
\vrule height7pt width0.5pt depth0.5pt \vrule height0pt width2pt
depth0pt} }

\begin{abstract}
General relativity postulates that the gravity field is defined on a Riemannian manifold. The field equations are $R^\mu_\nu = 0$ i.e. Ricci's curvature tensor vanishes. The field equations have to be augmented by natural physical requirements like orientability, time orientability and existence of a spinorial structure. Moreover, it is impossible to define the energy of the gravity field only by the metric tensor.
We suggest to impose an additional structure and consider parallelizable manifold i.e. manifolds for which a smooth field of frames exists. The derivation of the field equations is by an action principle. A Lagrangian, which is quadratic in the differentials, is defined. The minimum of the action is achieved at a $16 \times 16$ second order quasi linear system. It is of Laplacian type. The system admits a unique exact solution for a centrally symmetric, static and assimptotically flat field. The resulting metric is the celebrated Rosen metric, which is very close to the Schwarzchild metric. The two are intrinsicly different since the scalar curvature of the Schuarzschild metric is zero, in contrast to Rosen's which is positive. The suggested field admits black holes which are briefly discussed.
\end{abstract}
\newpage
\section{Basic framework.}
Einstein theory of gravity in it's pure case (without material fields) is defined on a differential manifold with a field of pseudo-Riemannian metric tensor. The field equations are 
\beqn
R^\mu_\nu = 0
\eeqn
The theory is very coherent and successful. Still it has shortcomings.
\begin{itemize}
\item Riemannian manifolds satisfying the field equations are short of a physical space-time. The manifold must satisfy some global topological conditions (orientability, time-orientability, existence of spinorial structure e.t.c.). 
\item The invariant energy of the gravity field can not be defined only by the metric tensor and it's derivatives. 
\item In Lagrangian formalism Hilbert's action is of a very special type. It contains the second order derivatives of the field variables (components of the metric tensor).This situation is not consistent with the other physical theories (electromagnetism).
\end{itemize} 
In order to deal with the difficulties above we suggest to study \it parallelizable manifolds\rm [Ca], [Do].  A parallelizable manifold is a differentiable manifold $M$, for which the frame bundle $FM$ allows a global cross-section. This type of a differentiable manifold satisfies the necessary topological conditions. Every Riemannian manifold is locally parallelizable. Geroch theorem [Ge] states: The parallelizability of the four dimensional Lorentzian manifold is a necessary and a sufficient condition for the existence of spinorial structure. \\
Let $\Phi^a \quad a=0,1,2,3$ be a local frame of differential 1 forms, smoothly depending on the point $x \in M$.   
Let the frame bundle $FM$ be equipped with a Hodge dual map. It is the $*$-operator acting on the monomials of the 1-forms  in the following manner 
\begin{equation}
*(\Phi^{a_0} \wedge ...\wedge \Phi^{a_k}) = (-1)^s \Phi^{a_{k+1}} \wedge ...\wedge \Phi^{a_n},
\end{equation}
where all indeces are different and are taken in such an order, that $\{a_0,...,a_{n}\}$  is an even permutation of $\{0,...,n\}$. $s$ is defined by
\begin{equation}
s=\left \{ \begin{array}{ll}
         0,& \textrm {if $ a_{j}=0 $ for $ 1 \le j  \le p$} \\
         1,& \textrm {if $ a_{j}=0$ for $p+1 \le j \le n $}
\end{array} \right.
\end{equation}
(By comparison, for elliptic geometry $s=0$).\\
There exists a unique metric so that prescribed frame is orthonormal with respect to it. The components of this metric are
\begin{equation}
g_{\mu\nu}=\eta _{ab} \Phi^a_\mu \Phi^b_\nu,
\end{equation}
where $\Phi^a_\mu$ are the components of the 1-form $\Phi^a$ in some local coordinate system  $x^\mu$
\begin{equation}
\Phi^a=\Phi^a_\mu dx^\mu,
\end{equation}
and $\eta _{ab}=diag(-1,1,1,1)$ is the Lorentzian metric tensor.\\
Thus the \it space-time \rm is defined as a triple $(M,\Phi^a,*)$, where $M$ is a four-dimensional $C^\infty$-differentiable manifold, $\Phi^a$ are four independent differential 1-forms and $*$ is the Hodge dual map. \\
Let $F$ be an algebra of real $C^\infty$ functions. Let $\Omega^p$ be the module of differential $p$-forms on $M$. 
Let $d$ be the exterior derivative $d: \Omega^p \to \Omega^{p+1}$.
Let $\delta=*d*:$ be the coderivative $\delta: \Omega^p \to \Omega^{p-1}.$

The Hodge-de Rham laplacian [Ho], [dR] is defined by  
\begin{equation}
\triangle=d\delta + \delta d.
\end{equation}
For any frame $\Phi^a$ the laplacian $\triangle \Phi^a$ is an invariant second order differential. \\
On a flat manifold $d\delta + \delta d$ is the ordinary laplacian, on functions it is the Laplace-Beltrami operator. \\
In [H-MC-M-N] a comprehensive theory of possible equations arising from an action principle for $(M,g_{\mu\nu},\Phi^\lambda,{\Gamma_\alpha}^\beta)$ is developed. There all the variables are considered to be independent. We deal only with gravity. Thus, in our theory, it is sufficient to take only the $\Phi^a$ as independent variable. The $g_{\mu,\nu}$ and ${\Gamma_\alpha}^\beta$ are defined by the $\Phi^a$.\\
We postulate that the field equations are 
$$\triangle \Phi^a + \lambda(x)\Phi^a=0$$
What is the motivation for taking the equations above? This is a proclamation about the space-time in which we live. Philosophically, this choice is as good as Einstein's $R^\mu_\nu=0$. It will be seen in the sequel that it is impossible to distinguish, by the latest experiment techniques, between Einstein theory and theory presented in this article.

\section{The variation principle and the field equation.}

Consider an action
\begin{equation}
S=\eta_{ab}\int_M d\Phi^a \wedge *d\Phi^b - \delta \Phi^a \wedge *\delta\Phi^b
\end{equation}
 The Lagrangian above is quadratic in the first order derivatives in conformity with classical physics. By comparison the Lagrangian for the action in General Relativity involves the second order derivatives. The variation of the functional is taken relative to variations of the 1-forms that commute with the star operator. It follows that the volume of the element $*1=\Phi^0 \wedge \Phi^1 \wedge \Phi^2 \wedge \Phi^3$ is preserved.\\
Let us consider an effective functional
\begin{equation}
 S_\lambda[\Phi^a] = \eta_{ab} \int_M[(d\Phi^a,d\Phi^b)-   (d*\Phi^a,d*\Phi^b)+\lambda(x)((\Phi^a,\Phi^b)-\eta^{ab})]*1
\end{equation}
where $\lambda(x)$ - is a Lagrange multiplier. 

The variation of the functional $$\delta S[\Phi^a]=S[\Phi^a+\delta \Phi^a]-S[\Phi^a]$$ takes the following form
$$\delta S[\Phi^a]= \eta_{ab}\int_M [(d\delta\Phi^a, d\Phi^b) +(d\Phi^a, d\delta\Phi^b) - (d\delta*\Phi^a, d*\Phi^b) -(d*\Phi^a, d\delta*\Phi^b) +$$
$$+\lambda(x)((\delta\Phi^a,\Phi^b)+(\Phi^a,\delta\Phi^b))+\delta\lambda(x)((\Phi^a,\Phi^b)-\eta^{ab})]*1=$$
$$=-\eta_{ab}\int_M[d\delta\Phi^a\wedge* d\Phi^b+d\Phi^a\wedge* d\delta\Phi^b-
d\delta*\Phi^a\wedge* d*\Phi^b-d*\Phi^a\wedge* d\delta*\Phi^b+$$
$$+\lambda(x)(\delta\Phi^a\wedge*\Phi^b+\Phi^a\wedge*\delta\Phi^b)-\delta\lambda(x)(\Phi^a\wedge*\Phi^b+\eta^{ab}*1)]$$
Using the Leibnitz rule for the exterior product and the symmetry of the expression, the first member can be rewritten as
$$d\delta\Phi^a\wedge*d\Phi^b = d(\delta\Phi^a\wedge*d\Phi^b)+\delta\Phi^a\wedge d*d\Phi^b$$ 
the second - $$d\Phi^a\wedge* d\delta\Phi^b=d\delta\Phi^b\wedge*d\Phi^a
=d(\delta\Phi^b\wedge*d\Phi^a)+\delta\Phi^b\wedge d*d\Phi^a =$$
$$=d(\delta\Phi^a\wedge*d\Phi^b)+\delta\Phi^a\wedge d*d\Phi^b $$ 
the third - $$d*\delta\Phi^a\wedge *d*\Phi^b=d(*\delta\Phi^a\wedge *d*\Phi^b) + *\delta\Phi^a\wedge d*d*\Phi^b=$$
$$=d(*\delta\Phi^a\wedge *d*\Phi^b) - \delta\Phi^a\wedge *d*d*\Phi^b$$
the fourth - $$d*\Phi^a\wedge *d*\delta\Phi^b=d*\delta\Phi^b\wedge *d*\Phi^a=d(*\delta\Phi^b\wedge *d*\Phi^a)+*\delta\Phi^b\wedge d*d*\Phi^a=$$
$$=d(*\delta\Phi^b\wedge *d*\Phi^a)-\delta\Phi^a\wedge *d*d*\Phi^b$$
and the last one $$\lambda\Phi^a \wedge *\delta\Phi^b=\lambda\delta\Phi^b\wedge*\Phi^a=\lambda\delta\Phi^a\wedge*\Phi^b$$
Omitting total differentials, this expression can be rewritten as
$$\delta S[\Phi^a]= 2\eta_{ab}\int_M \delta\Phi^a\wedge[d*d\Phi^b+*d*d*\Phi^b+\lambda(x)*\Phi^b]+\delta\lambda(x)(\Phi^a\wedge*\Phi^b+\eta^{ab}*1)$$
Due to the arbitrariness of the variation one gets the Lagrange-Euler equations for the action
\begin{equation}
\triangle\Phi^a +\lambda(x)\Phi^a=0
\end{equation}
The contraction of the equation gives
\begin{equation}
B^a_b-{1\over 4} B\delta^a_b=0,
\end{equation}
where the matrix $B^a_b$ is defined by the expression
$$\triangle \Phi^a=B^a_b \Phi^b$$
and $B$ is the trace of this matrix.
\section{Static central symmetric solution of the equation.}
 Using the Einstein-Mayer result [E-M] the general spherical symmetric static frame can be written in the following form
\begin{equation}
\Phi^0 = f_1(s) dx^0
\end{equation}
\begin{equation}
\Phi^i = f_2(s) dx^i  + g(s) x^i dx^0
\end{equation}
where $s = x^2 + y^2 + z^2$. The field equations (10) for this fundamental form, comprise an  over-determined system of four ordinary differential equations for the three functions $f_1(s)$,
$f_2(s)$,$g(s)$. The only possibility to satisfy the o.d.e. system is to take $$g(s) = 0$$
Thus, $\Phi^a$ can be expressed as
\begin{equation}
\Phi^0 = e^{\mu(s)} dx^0
\end{equation}
\begin{equation}
\Phi^i = e^{\lambda(s)}dx^i
\end{equation}

By a direct computation of $d\delta+\delta d$ the Hodge-de Rham Laplacian on this fundamental form can be written as
\begin{equation}
\triangle \Phi^0 = -2e^{-2\lambda}[3\mu^\prime + 2s(\mu ^{\prime \prime} +\mu^ \prime \lambda^ \prime)] \Phi^0
\end{equation}
$$
\triangle \Phi^1 = -2e^{-2\lambda}[\mu^\prime +4\lambda^\prime+2x^2(\mu^{\prime \prime} +2\lambda^ {\prime \prime} - \lambda^\prime \mu^\prime -2 {\lambda ^\prime}^2) + 2(y^2+z^2)(\lambda^{\prime \prime }+ \mu^\prime \lambda^\prime] \Phi^1 -$$
\begin{equation}
-4 e^{-2\lambda}(\mu^{\prime \prime}+\lambda^{\prime \prime}-2\mu^\prime \lambda^\prime -2{\lambda^\prime}^2)(xy\Phi^2+xz\Phi^3)
\end{equation}
Thus by (10) for $a\not = b$
$$\mu^{\prime \prime}+\lambda^{\prime \prime}-2\lambda^\prime (\mu^\prime +\lambda^\prime) = 0 $$
The first integral of which is
 $$\mu^\prime +\lambda^\prime = C e^{2\lambda}$$
Being interested only in asymptotically-flat solutions, we take 
 $$\mu^\prime +\lambda^\prime = 0$$
Using this relation, equation (10) can be written  for $a = b$ as
$$2s\lambda^{\prime \prime}+3\lambda^\prime = 0$$
Integration of this equation results in
$$ \lambda = C_1 +{C_2\over \sqrt s}$$
consequently
$$ \mu = C_1 -{C_2\over \sqrt s}$$ 
Omitting the constants $C_1,C_3$ as a scale factor and taking $C_2= -m$ one gets a solution 
\begin{equation}
\Phi^0=e^{-{m\over r}} dt  \qquad \Phi^i=e^{m\over r} dx^i
\end{equation}
The metric tensor can be written now in the following form
\begin{equation}
ds^2 = e^{-2{m\over r}} dt^2 - e^{2m\over r} (dx^2+dy^2+dz^2)
\end{equation}
This is the Rosen metric [Ro]. The result above can be restated in the form of a Birkhoff-type theorem.

\pro The gravity field equations have a unique, static, spherical symmetric, asymptotically-flat solution.
\epro
The Tailor expansion of the line element (1) up to and including the order $1\over r^2$ takes the form
\begin{equation}
ds^2=(1-{{2m}\over{ r}} + {{2m^2} \over {r^2}}+...) dt^2 - ( 1+{{2m} \over {r}} + {{2m^2} \over {r^2}}+...) (dx^2+dy^2+dz^2)
\end{equation}
The Schwarzchild line element, in the isotropic coordinates, is
\begin{equation}
ds^2=[{1-{m\over{2r}}\over {1+{2m \over{2r}}}}]^2dt^2-(1+{m\over{2r}})^4(dx^2+dy^2+dz^2)
\end{equation}
and it's Tailor expansion up to the same order is
\begin{equation}
ds^2=(1-{2m\over r} + {{2m^2} \over {r^2}}+...)dt^2 - (1+{2m \over r} +{3\over 2} {{m^2} \over {r^2}}+...)(dx^2+dy^2+dz^2)
\end{equation}
The difference between these two line elements is only in the second order term of the spatial part and in the third order term of the temporal part of the metric.\\
It is easy to see that the Rosen metric is essentially different from the Schwarzchild metric. Indeed, the scalar curvature is
\begin{equation}
R=2{m^2\over r^4}e^{-2{m\over r}} \ne 0
\end{equation}
Thus, there exists no coordinate system that transforms Schwarzchild solution to the solution above.
\section{Physical effects and black holes.}
 Let us recall the Geodesic Principle. By this principle, the trajectory of a test particle in a static centrally symmetric field is a geodesic on the manifold, defined via the metric connection.
It turns out that the experimental tests of General Relativity and the Geodesic Principle are based on the first two terms of the temporal component and the first term of the spatial component.
For our theory in it's present form, the Geodesic Principle has to be used as well. Schwarzschild and Rosen metrics agree to the second order in the temporal components and to the first order terms in the spartial part.
Thus within the framework of the restricted two body problem the two metrics are experimentally indistinguishable, even by the current measurements.
 Schwarzschild line element admits a distinguished sphere with ``Schwarzschild radius'' $r=2m$. Kruskal showed how to continue the metric, smoothly, into the ball $r<2m$. Nevertheless, a light ray can not escape from the ball. This is a black hole .\\
By comparison, Rosen metric is regular up to the origin. In particular, there is no ``Schwarzschild radius''. Black holes, however, do exist. \\
The geodesic line defined by Rosen metric is planer. Take it, in polar coordinates to be defined by $\theta=\pi/2$. Denote $u=1/r$ and $u^{\prime}={{du}\over {d\phi}}$. Then for Rosen metric the null-geodesics satisfy
\begin{equation}
u^{\prime\prime}+u=2m(u^{\prime2}+u^2).
\end{equation}
This equation is exact. \\
By comparison, the corresponding equation for Schwarzschild line element, in isotropic coordinates is:
\begin{equation}
u^{\prime\prime}+u=[2m-m^2/2u+O(m^3u^2)](u^{\prime2}+u^2).
\end{equation}
(When $u$ is large one may not discard $m^2/2u$ and the higher order terms.)\\
Let us consider the Rosen metric.
If $u>(2m)^{-1}$ (i.e. $r<2m$) and $u^{\prime}\ge0$ then $u^{\prime\prime}>0$. Therefore, $u$ increases indefinitely. Hence, every incoming ray that hits the ball is trapped. \\
Consider now outgoing rays.
By the equation above if $2mu > 1$ then $u^{\prime\prime} > 0$. \\
If, for some point on the trajectory, $u^{\prime}=0$ while  $2mu > 1$ (i.e. the trajectory is still in the ball) the light ray, again, will be trapped. As the point of emanation of the outgoing light ray gets closer to the origin the cone from which it will escape becomes very narrow. Let us furnish a sufficient condition.
\pro If at the point of emanation of the outgoing light ray $$u \ge 6sm^{-1},$$ where $s \ge 2$ while $$u^{\prime}=-m^{-1}s^2$$ then the light ray is trapped
\epro
\prf
By induction. Let $I$ (in the angle $\phi$) be the interval for which $$-s^2m^{-1} \le u^ {\prime} \le-(s-1)^2m^{-1}.$$ 
On that interval 
$$u^{\prime\prime} \ge 2m^{-1}(s-1)^4.$$
The length of $I$ is smaller than ${s^2-(s-1)^2} \over {2(s-1)^4}$. The extremal case is $u = 6sm^{-1}$. Then on $I$
 $$u \ge 6sm^{-1}-s^2m^{-1}{{s^2-(s-1)^2} \over {2(s-1)^4}} \ge 6(s-1)m^{-1}.$$
Since $$u^{\prime} \le -(s-1)^2m^{-1}$$
we may take
$$u \ge 6(s-1)m^{-1} \qquad  \qquad u^ {\prime} \le -(s-1)^2m^{-1}$$ 
one may reduce the problem and take 
$$u^{\prime} = -(s-1)^2m^{-1}.$$
Thus, by induction, the problem is reduced to 
$$u^{\prime}=4m^{-1}, \qquad u \ge 6m^{-1}  \qquad (\text {i.e.} \quad s=2).$$
By the argument above, it is clear that $u={1\over 2} m^{-1}$ can not be attained. \\
Thus the cone from which a light ray can escape is quite narrow. It includes the radial direction. There $u^{\prime}=\infty$.\\
\eprf

\centerline{\bf References}

{[Ca]} Cartan, E., {\it Geometry of Riemannian spaces}, Math. Sci. Press, New York, 1983, p.217-231.

{[dR]} de Rham, G., {\it Varietes differentiables}, Herman,Paris,1973, p.117-135.

{[Do]} Dodson, C.T.J., {\it Categories, bundles and spacetime topology.}, Shiva Mathematics series; 1, 1980.

{[E-M]} Einstein, A., Mayer, W.,{\it Zwei strenge statische L\"osungen der Feldgleichungen der einheitlichen Feldtheorie}, Sitzungsber. preuss, Acad. Wiss., phys.-math.,K 1, 1930, 110-120.

{[H-MC-M-N]} Hehl, F.W., McCrea, J.D., Mielke, E.W., Ne'eman, Y., {\it Metric-affine gauge theory of gravity}, Phys. Rept.,258, 1995, 1.
 
{[Ho]} Hodge, W.V.D., {\it The theory and applications of harmonic 
integrals}, Cambridge University Press,1989, p.103-111. 

{[Ge]} Geroch, R.P., {\it Spinor structures of space-time in general relativity I}, J. Math. Phys.,1968,9, 11, 1739-1744.

{[Ro]} Rosen, N., {\it A bi-metric theory of gravitation}, General Relativity and Gravitation, 1973, 4, 6, 435-447.

\end{document}